\documentclass[manuscript]{acmart}
\AtBeginDocument{%
  }

\usepackage{xcolor}
\usepackage{tcolorbox}
\usepackage{multirow}
\usepackage{xr}
\usepackage{graphicx}
\usepackage{subcaption} 
\usepackage{placeins}
\usepackage{enumitem}
\usepackage{tabularx}
\usepackage{enumitem}

\newcommand{\recsection}[3]{
    \paragraph{\textit{#1}} {#2}
    \ifthenelse{\equal{#3}{}}{}{\newline\hspace*{2em}\textit{Consider}: #3}
}

\begin{document}

\title[Early-Stage GenAI Support for Novice Artists]{Exploring Opportunities to Support Novice Visual Artists' Inspiration and Ideation with Generative AI}

\author{Cindy Peng\textsuperscript{*}}
\email{}
\affiliation{%
  \institution{Carnegie Mellon University}
  \country{USA}
  \authornote{Co-first authors contributed equally to this research.}
}

\author{Alice Qian\textsuperscript{*}}
\email{aqzhang@andrew.cmu.edu}
\affiliation{%
  \institution{Carnegie Mellon University}
  \country{USA}
}

\author{Linghao Jin}
\email{linghaoj@usc.edu}
\affiliation{
    \institution{University of Southern California}
    \country{USA}
}

\author{Jieneng Chen}
\email{jchen293@jhu.edu}
\affiliation{
    \institution{Johns Hopkins University}
    \country{USA}
}
\author{Evans Xu Han}
\affiliation{
    \institution{Stanford University}
    \country{USA}
}

\author{Paul Pu Liang}
\email{ppliang@mit.edu}
\affiliation{
    \institution{Massachussetts Insitute of Technology}
    \country{USA}
}

\author{Hong Shen}
\email{hongs@andrew.cmu.edu}
\affiliation{%
  \institution{Carnegie Mellon University}
  \country{USA}
  }

\author{Haiyi Zhu}
\email{jhsieh2@andrew.cmu.edu}
\affiliation{
   \institution{Carnegie Mellon University}
  \country{USA}
}

\author{Jane Hsieh}
\email{jhsieh2@andrew.cmu.edu}
\affiliation{
   \institution{Carnegie Mellon University}
  \country{USA}
}

\renewcommand{\shortauthors}{Cindy Peng, et al.}

\begin{abstract}
Recent generative AI advances present new possibilities for supporting visual art creation, but how such promise might assist novice artists during early-stage processes requires investigation. How novices adopt or resist these tools can shift the relationship between the art community and generative systems. We interviewed 13 artists to uncover needs in key dimensions during early stages of creation: (1) quicker and better access to references, (2) visualizations of reference combinations, (3) external artistic feedback, and (4) personalized support to learn new techniques and styles. Mapping such needs to state-of-the-art open-sourced advances, we developed a set of six interactive prototypes to expose emerging capabilities to novice artists. Afterward, we conducted co-design workshops with 13 novice visual artists through which artists articulated requirements and tensions for artist-centered AI development. Our work reveals opportunities to design novice-targeted tools that foreground artists' needs, offering alternative visions for generative AI to serve visual creativity.
\end{abstract}




\maketitle

\section{Introduction}
Visual art creation is a complex, multistage process requiring significant time, effort, and material investment. Recent advances and uses of generative AI show potential for supporting this intensive creation process through consistent, guided generation and editing \cite{zhang2023text} of increasingly realistic and detailed visual outputs \cite{podell2024sdxl}. 
Such progress, coupled with lowered barriers of access, sparks excitement around the potential for visual generative models to revolutionize and democratize art education and expand creative participation \cite{paradox, democratize, edu}. 

While many creativity support tools enabled by visual generative AI have explored ways to augment the workflows of professional artists and designers across various domains \cite{wang2025aideation, GenQuery, architectural}, novice visual artists often engage in art-making to explore new capabilities \cite{clinical, personal}, practice self-expression \cite{real_benefits} or challenge conventional aesthetic and creative boundaries \cite{decline_novelty, art_implications}. For those users, the general-purpose and zero-shot nature of recent models offers an exciting promise: to make art-making accessible to audiences with limited training \cite{lgtm}. In theory, such tools could empower more people to develop skills, enjoy the process of art making, and gain a deeper appreciation for visual craft \cite{jackson}. 

Yet, existing generative AI systems often fall short in supporting novice visual artists, particularly in the early stage of the creative process. Prior work has highlighted the potential of AI to augment early ideation \cite{art_implications, poet, lamuse}, but novice artists often find these tools not being effectively integrated into their creative processes \cite{unstraighten}, leaving them stressed and disoriented \cite{kawakami2024impact, lgtm}. 
Meanwhile, artists remain deeply distrusting of generative systems due to harmful downstream impacts on members of the artistic community (e.g., economic and reputational harms arising from inadequate attribution or compensation), raising questions about what artist-centered, ethically sound uses and developments of generative AI tools could look like -- especially for those who just beginning their creative journeys. 

To date, much of the HCI and UIST literature has focused on understanding and supporting final-stage outputs
of \textit{professional} designers and artists \cite{design_professionals, shi2023understanding}, while generative AI support for the early stages of novice visual artists workflows remain underexplored -- overlooking a key opportunity to bridge the gap between creativity support tools and emerging members of the artistic community. In particular, few studies have examined how novice artists \cite{hu2025designing} adopted co-creation and inspiration tools to meaningfully organize and reflect on inputs \cite{lamuse}, leaving a critical gap in understanding how emerging generative capabilities may align with novice visual artists' \textbf{early-stage workflows} involving practices that start at the definition of a piece and ending with the use of references for revisions to pieces  \cite{botella2018stages}.

To address these gaps, this study investigates how generative AI can meaningfully support the early-stage workflows of novice visual artists. 
We employed a two-step approach: (1) We interviewed 15 novice artists to understand their existing practices, initial perceptions of available generative tools, as well as painpoints to their artistic creation processes (\S\ref{interview}); next (2) We mapped these expressed needs to state-of-the-art technical capabilities and develop six functional prototypes, which we evaluated in co-design workshops with 14 novice artists (\S\ref{codesign}).
Leveraging state-of-the-art capabilities (e.g., medium and style transfer, visual combinations and editing, step-by-step tutorials, and extensions of 2D into interactive 3D viewing),
these interactive prototypes allowed us to probe novice artists for reactions and design objectives they prioritized for future forms of tooling support.

We ask the following research questions:
\begin{enumerate}
\item[\textbf{RQ 1}] 
\textbf{What processes do novice visual artists undertake to navigate early stages of their creative workflows, and what challenges do they encounter?}
\item[\textbf{RQ 2}] 
\textbf{How do novice visual artists perceive the potential of emerging generative AI capabilities to support their early-stage needs, and what tensions do they foresee around their integration?}

\end{enumerate}


Building on the above questions, we map early-stage practice, probe emerging capabilities through co-design and prototypes, and distill implications for toolmakers. Accordingly, this paper contributes.
\begin{itemize}
     \item An empirical account of how novice artists use tools in the early stage and the challenges they face.
    \item Empirically grounded capability requirements for early stage support that artists want in tools.
    \item A synthesis of the tensions these tools raise and the implications for design and policy.
\end{itemize}
\section{Related Work}
Below, we first overview theory and empircal findings around the artistic creation process and steps, followed by existing creativity support tools for visual artists and designers. Finally, we summarize literature capturing artistic community perceptions and concerns around generative AI support tools, and compare these with recent excitement for around more democratized opportunities for artistic creation to surface a key gap for supporting early-stage workflows of novice artists, which we aim to address in this study.

\subsection{Creative Workflows of (Professional) Visual Artists}
Visual art creation is complex, iterative and resource-intensive process, containing subprocesses and tasks that vary widely depending on the artists' expertise, intentions, as well as the choice and availability of materials, medium and aesthetics \cite{multivariate, subproccess}. Since an entire century ago, psychology scholars sought to model the stages and dimensions of the artistic creation process:
\citet{wallas} proposed the four macroprocesses of \textit{preparation}, \textit{incubation}, \textit{illumination}, and \textit{revision} \cite{wallas, wallas_followup}, 
\citet{soi} introduced \textit{divergent} and \textit{convergent} thinking, \citet{vision} identified the \textit{problem formulation} and \textit{solution} stages (pre-drawing versus during drawing and feedback), while \citet{geneplore} separated the creative activity into the \textit{generate} and \textit{explore} processes. 

More empirically, \citet{modeling} offered the first grounded macro-process model of professional artmaking, which consisted of (1) \textit{artwork conception} (2) \textit{idea development} (3) \textit{making of the artwork} and (4) \textit{finishing the artwork}. The first step of \textit{conception} also implicates (sub)processes of orientation \cite{osborn}, \textit{preparation} \cite{wallas}, \textit{concentration} \cite{ecological} and \textit{a goal of creation} \cite{furst}. The second step (\textit{idea development}) --- wherein the artist (re)structures the idea --- is highly tied to the relaxed and subconscious idea-association process of \textit{incubation} \cite{wallas} as well as \textit{intimation} \cite{wallas_followup}, which lead to \textit{insight} -- the illuminating moment when an idea or image emerges, a process that requires prior thinking \cite{boden}. Then, the \textit{(3) making of the artwork} describes where the idea is realized. Finally, the \textit{(4) finalization} step or \textit{validation} occurs -- where the art might be deemed complete, get elaborated on, or even abandoned. To better situate artistic process in the creator's life and contexts, \citet{multivariate} took a more action-oriented framework to qualitatively reconstruct professional workflows, finding additional stages including the \textit{documentation and reflection} phase between \textit{(1) artwork conception} and \textit{(2) idea development} where the artists gathers information about required materials and technologies, as well as the refinement of \textit{(3) making of the artwork} into subprocesses of \textit{first sketches} and the \textit{testing of forms and ideas} that origniate from reflection.

But while professional artmaking has been extensively examined, the literature around student (or novice) artist practices remains relatively sparse --- leaving their wandering, nonlinear and unsystematic approaches underexplored \cite{ecological, mfa, modification}. The handful exceptions offered (1) insights that novices engage relatively more with earlier stages of the professional practice --- e.g., idea conception, documentation and development, discovery-oriented behaviors \cite{solving} --- since they are still learning how to produce and evaluate ideas and (2) an inventory of 17 stages of novice artists' practices, where key early processes included immersion, reflection, inspiration, research, ideation and selection \cite{botella2018stages}. Despite such detailed inventory of cognitive microprocesses, there remains a gap in understanding how novice artists interact with digital mediums \cite{cocreative} or technology support during their creative processes, which this study aims to addresss.

\subsection{Co-Creative AI-Support Tools for Early-Stage Artmaking}

Meanwhile, studies at the intersection of the CHI and UIST communities showcase various forms of digital creativity support tools \cite{centric, PortraitSketch, k_sketch}, some of which even center novice artists \cite{bob} while others adapt generative AI systems for co-creation~\cite{wang2025aideation, o2015designscape, ko2023large, fan2024contextcam, Midjourney, AdobeFirefly, everybody_sketch}.
Following the taxonomy of multimodal image-text generative models~\cite{liang2024foundations}, we categorize these co-creative AI systems into (1) summarization tools that reduce information content by highlighting salient parts of inputs, (2) translation methods that modify or edit information along dimensions while keeping other desired attributes consistent, and (3) creation tools that generate data and increase information content while maintaining coherence and controllability.

\textit{Summarization}-based approaches highlight specific parts of visual inputs ~\citep{kong2019understanding}, strip away details to obtain simpler sketches~\citep{swiftsketch}, or summarize the image using key attributes or database references~\citep{sermuga2021uisketch}. Such tools support (1) data visualization in storytelling and presentation~\citep{kong2019understanding}, indexing and retrieving relevant user interfaces for designers~\citep{sermuga2021uisketch}, as well as in summarizing visual interfaces into short language phrases for easier user understanding~\citep{wang2021screen2words}. Summarization-based approaches also decompose visual content into multiple sub-parts including step-by-step tutorials~\citep{PaintsUNDO} or stroke references~\citep{neutralstroke}, which can support artists in learning techniques for achieving specific effects, in addition to making the creative process more transparent. By highlighting key steps, summarization-based approaches also afford users
fine-grained control over the drawing process~\citep{paintsalter}.

\textit{Translation} approaches for co-creation can be performed using retrieval, generation, or editing. Retrieval-based methods return relevant visual information in a manner that is semantic similar, such as retrieving images based on textual descriptions~\cite{GenQuery, VSC}, multiple references~\citep{OrganizingPhoto}, or based on target attributes such as color, material, and style~\citep{wu2025qwen, gpt4v, liu2024llavanext}. Generative versions have also been proposed, particularly in developing more interactive text-to-image generative models that automatically refine text prompts to improve user satisfaction~\citep{wang2024promptcharm} and diversify the output space of generated images for personalization~\citep{poet}. Editing-based approaches seek to maintain a portion of the information, such as visual styles, while making targeted changes to the other parts like content or backgrounds. Recent advances in generative AI have enabled fine-grained style transfer and medium adaptation through lightweight model tuning techniques~\citep{blora, flux}, perspective or background changes with multimodal editing tools~\citep{wu2025qwen, backgroundremove}. A related line of work explores visual editing through direct manipulation, allowing designers to annotate or draw on an image and supplement these edits with text descriptions, thereby reducing reliance on purely textual prompts and giving users greater agency over the generative process~\cite{liu2025magicquill, ati}.

Finally, \textit{creation} extends the information in a consistent way, often based on targeted sampling from generative models. For example, recent work has investigated how multiple elements from different source images or sketches can be combined in a consistent manner to produce new compositions~\cite{magiccolor, yan2025imagereferencedsketchcolorization}. Other tools convert 2D inputs into 3D assets for spatial exploration across multiple views in a geometrically-consistent manner~\citep{trellis, wang2024gaussianeditor, li2025voxhammer, brooks2023instructpix2pix}. Finally, image-to-video generative models help convert a web page into short videos to support designers in marketing~\citep{chi2020automatic} and can convert documents into narrated instructional videos for education~\citep{chi2022synthesis}.










Collectively, these systems highlight the growing potential of AI tools to not only generate and retrieve but also summarize, recombine, and transform references in ways that support iterative and exploratory workflows. 
However, much of this work centers on professional designers and commercial outcomes rather than artistic self-expression or learning. Our work addresses this gap by focusing on the early-stage inspiration workflows of novice artists to explore how AI can serve as a partner in their creative exploration, instead of as a tool to produce finished works.

\subsection{Artists' Perceptions of Generative AI versus Visions of Democratization and Access}

Such recent advances spurs a wave of excitement for the transformative potential of generative AI for democratizing access to the artmaking process. \citet{chatterjee} outlines its power potential to produce emotional and evocative works, as well as to challenge traditional notions of aesthetics -- thereby democratizing art ``by making the tools of artistic creation accessible to a broader audience'' \cite{liberation}. \citet{imperial} argues that generative AI have been facilitated to serve as a new form of every day life, extending the \textit{imperical mode of living} into the sphere of everyday aesthetic production. 
  
But despite the potential for generative AI to support artistic co-creation, the artistic community remains skeptical and threatened by their potential to plagiarize \citep{shi2023understanding}, harm rights over attribution and ownership \citep{lima} as well as unfair models of compensation~\citep{kawakami2024impact, foreground} --- prompting calls for more inclusive and decolonial futures of AI in the arts~\citep{decolonial}.
~\citet{giacomin2023intersection} highlighted contested debates surrounding copyright issues, ethical and legal concerns, the impact on the art industry, questions of authorship and authenticity as well as how AI reshapes our understanding of creativity and art itself.
~\citet{decline_novelty} found peak artwork content novelty to increase over time but average novelty to decline, suggesting an expanding but inefficient idea space, which calls for more harmonious blending of human exploration and AI exploitation. 
Finally,~\citet{unstraighten} found queer artists to struggle in using these models due to various embedded normative values, such as hyper-positivity and anti-sexuality.

A key factor impacting artists' perception of generative AI is their level of experience. Experienced visual professionals with more confidence in their originality, creativity, and empathic skills were more likely to find generative AI's role as assistive~\citep{li2024user, kawakami2024impact}. 
Meanwhile, skill degradation, job replacement, and creativity exhaustion adversely impact junior workflows~\citep{li2024user}.
\citet{epstein2023art} proposes that disruptive technological advances will not necessarily end all human artistic creation, but rather produce more complex effects and shifts in aesthetics, roles and practices. ~\citet{ko2023large} and~\citet{inie2023designing} both found significant potential and new conceptions on what constitutes creativity in relation to generative AI --- acknowledging AI's versatile roles with high usability to support creative works in automating the creation process (i.e., automation), expanding their ideas (i.e., exploration), and facilitating or arbitrating communication (i.e., mediation). But despite such visions to support the artistic creation process, the current body of work falls short in understanding how such digital and generative forms of support align with grounded needs of \textit{novice} artists \cite{student_ai}, who represent a much broader segment of the artistic community than professional or commercial practitioners. Our work strives to not only characterize ways that novice artists currently perceive and use generative support in their workflows, but to also understand how their common challenges may align with emerging technical capabilities.

\FloatBarrier
\begin{figure}[t]
    \centering
    \resizebox{.7\textwidth}{!}
    {\includegraphics{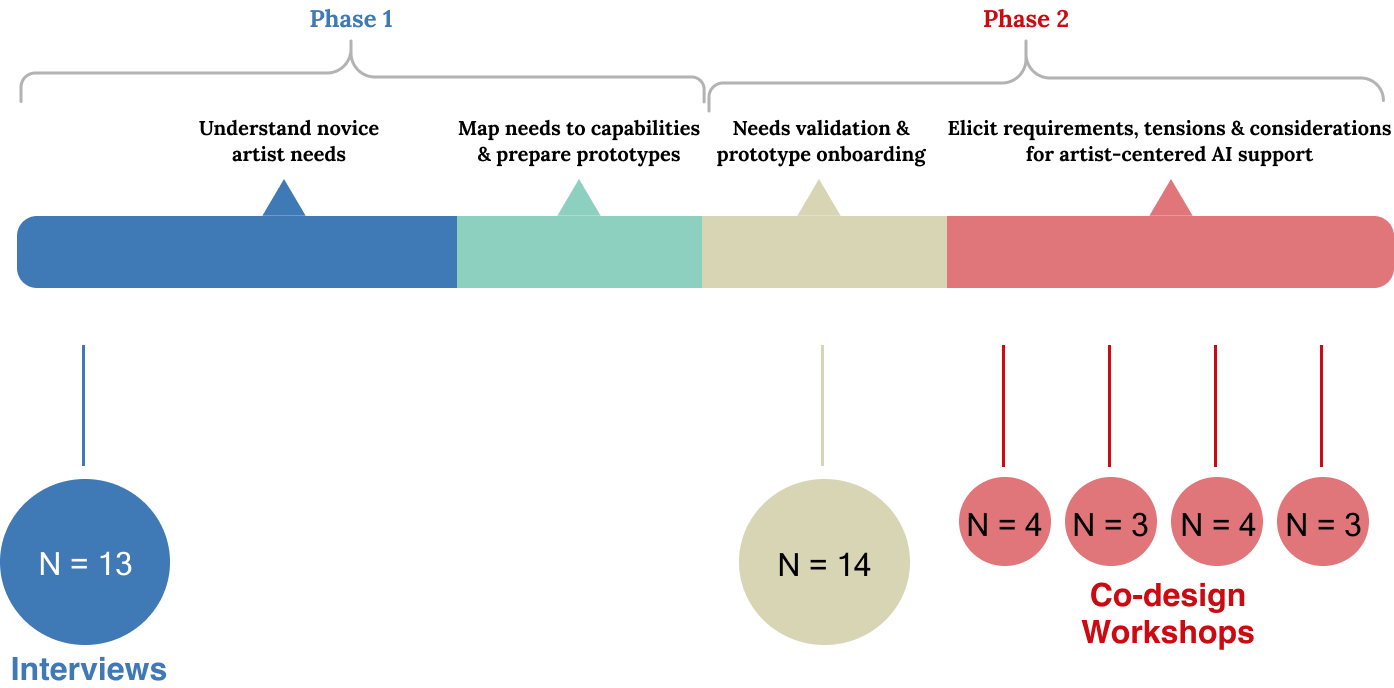}}
    \caption{Overview of Phase 1 \& Phase 2 Methods.}
    \label{fig:method}
\end{figure}

\section{Methods}
We start by describing and reflecting on our positions as researchers in this project. Then we describe our process conducting semi-structured interviews with novice artists (\S\ref{interview}), resulting in a taxonomy of their challenges, which we mapped to emergent generative capabilities (\S\ref{map}). Between the two phases, our process
mapping artistic challenges and needs to nascent techniques helped us identify relevant tools and build prototypes (See \autoref{fig:prototypes}) that we presented to participants as probes in Phase 2 (\S\ref{probes}).
This two-phase approach grounded our research in the lived realities and vocabularies of the artistic population we sought to understand and support.

\subsection{Phase 1: Interview Methods} \label{interview} 
To understand nuanced perspectives of novice artists' needs, we conducted semi-structured qualitative interviews with 13 novice artists working across a range of mediums and practices. We first asked participants to describe the early-stages of their art-making processes from the inspiration that leads to the conception of an idea for a piece to later searching for references to use. We then asked follow-up questions about how participants encounter, search for, and get references and inspiration. We also asked participants about the challenges they experienced using digital and AI tools. This approach allowed us to capture rich, situated insights about the needs artists have in the early stages of visual art creation. Our interviews focused on participants' current practices with visual art creation in the early stages, tools used, and challenges faced. This initial interview phase informed the development of a taxonomy of four key areas of support for novice artists.

\begin{table}
\centering
\small
\setlength{\tabcolsep}{3pt}
\begin{tabular}{@{}c p{3.0cm} p{2.5cm} p{2cm} p{5.0cm}@{}}
\toprule
ID & Physical mediums & Digital mediums & AI tools used & Reference sources \\
\midrule
P1 & Watercolor, Charcoal, Oil Painting & Animation & None & Google Image Search, Pinterest, Taking my own photos \\
P2 & None & Digital drawing painting & NA & Google Image Search, Taking my own photos \\
P3 & Watercolor & None & None & Google Image Search, Pinterest, Artist Reference Websites (e.g., ArtStation), Unsplash, Taking my own photos \\
P4 & None & None & None & Google Image Search, Pinterest, Taking my own photos \\
P5 & None & Digital drawing painting & None & Instagram \\
P6 & Pencil, Ink, Paint & Digital drawing painting & Midjourney & NA \\
P7 & Ink, Watercolor & Digital drawing painting & ChatGPT, DALL E, Other, Gemini & Taking my own photos \\
P8 & Pencil, Acrylic Paint, Other & Digital drawing painting, Digital art ipad & None & Tiktok \\
P9 & Pencil, Ink, Oil Paint & None & None & Instagram \\
P10 & Pencil, Ink, Watercolor & Digital drawing painting & None & NA \\
P11 & Pencil, Ink, Acrylic Paint, Pens & Digital drawing painting & None & ArtStation \\
P12 & Pencil, Acrylic Paint, Pastels, Watercolor, Chalk & Digital drawing painting & ChatGPT, DALL E, Midjourney, Stable Diffusion & Google Image Search, Pinterest, Taking my own photos \\
P13 & Pencil, Acrylic Paint, Pastels, gouache or acrylic gouache, embroidery, knitting & Digital drawing and painting & None &  Google Image Search\\
\bottomrule
\end{tabular}
\caption{Phase 1 participant backgrounds.}
\label{tab:interview-demographics}
\end{table}

\begin{table}
\centering
\small
\setlength{\tabcolsep}{3pt}
\begin{tabular}{@{}c p{3.2cm} p{3.2cm} p{2.5cm} p{5cm}@{}}
\toprule
ID & Physical media & Digital media & AI tools & Reference tools \\
\midrule
W1 & None & Digital drawing painting & NA & Google Image Search, Taking my own photos \\
W2 & Digital drawing painting & Adobe Photoshop, Adobe Illustrator, Other &  None & Pinterest, Artist Reference Websites (e.g., ArtStation), Taking my own photos \\
W3 & None & Digital drawing painting &  None & Instagram \\
W4 & Pencil, Ink, Acrylic Paint, Pens & Digital drawing painting &  None & ArtStation \\
W5 & Pencil, Ink, Watercolor & Digital drawing painting &  None & NA \\
W6 &  & None &  None & Google Image Search, Pinterest, Taking my own photos \\
W7 & Pencil, Acrylic Paint, Pastels, gouache/acrylic gouache, embroidery, knitting &  &  None & Google Image Search \\
W8 & Pencil, Ink, Paint & Digital drawing painting & Midjourney & NA \\
W9 & Pencil, Acrylic Paint, Other & Digital drawing painting &  None & Tiktok \\
W10 & Digital painting, Digital illustration & Procreate &  None & Google Image Search, Pinterest \\
W11 & pencil drawing 15 years, acrylic painting 10 years, watercolor painting 7 years; digital drawing 4 years & Pencil, Ink, Acrylic Paint, Watercolor & ChatGPT, Stable Diffusion & Google Image Search, Pinterest, Taking my own photos \\
W12 & Pencil, Ink & Digital painting & ChatGPT, DALL E, Midjourney, Stable Diffusion, Gemini, Claude & Google Image Search, Pinterest, Artist Reference Websites (e.g., ArtStation), Taking my own photos \\
W13 & Digital Art 6 years, pencil drawing 8 years, acrylic painting 3 years & Pencil, Acrylic Paint &  None & Google Image Search, Pinterest, Artist Reference Websites (e.g., ArtStation), Taking my own photos \\
\bottomrule
\end{tabular}
\caption{Participant backgrounds for media, AI usage, and references.}
\label{tab:workershop-demographics}
\end{table}

\subsubsection{Interview Recruitment.}
We recruited participants who self-identified as novice visual artists to better understand and support their early-stage creative practices. To ensure a diverse range of experiences, we sampled participants with varying degrees of familiarity with physical and digital art mediums. Recruitment was conducted through a combination of snowball sampling and calls posted on university-affiliated online platforms and listservs. All participants were based in the United States and fluent in English. See Table \ref{tab:interview-demographics} for participant demographics. Participants received \$30 USD for completing a 60-minute interview with at least one of the first 4 authors. All compensation was given via the form of online gift cards.

\subsubsection{Data Analysis.}
We conducted reflexive thematic analysis~\cite{braun2019reflecting} on interview transcripts to identify patterns in participants’ experiences and design feedback. Our process began with first-cycle open coding to surface recurring concepts across interviews and workshops. In the second cycle, we grouped codes into broader thematic categories that reflected early-stage creative processes and challenges that novice visual artists encounter. 
Data from the co-design sessions were analyzed separately with a focus on surfacing emergent design goals, constraints, and values articulated by participants. 
Three researchers collaboratively coded the dataset, engaging in memo writing throughout the process to document analytic decisions and emergent insights. Through iterative discussion and refinement, the coding scheme was revised to ensure consistency and alignment with the evolving thematic structure.

\subsection{Mapping Artistic Challenges to Generative Capabilities \& Developing Probing Prototypes}
\label{map}

The ground-up need-finding and analysis process resulted in Figure~\ref{fig:prototypes}, containing four higher-level and ten more granular challenges. Our research team worked together to categorize these under early stages of the novice artistic workflow, and mapped a selection of these with emerging generative techniques that our team perceived as potential matches. Based on identified matches, we selected six recently-released and open-sourced techniques that carry capabilities relevant to our participants' artistic needs, and subsequently built interfaces to surface these tools -- detailed in \S\ref{probes}. We note that we did not provide prototypes for all of the challenges because of readily available technologies (e.g., Google Search for semantic search). 

\begin{figure}[tbp]
    \centering
    \resizebox{\textwidth}{!}
    {\includegraphics{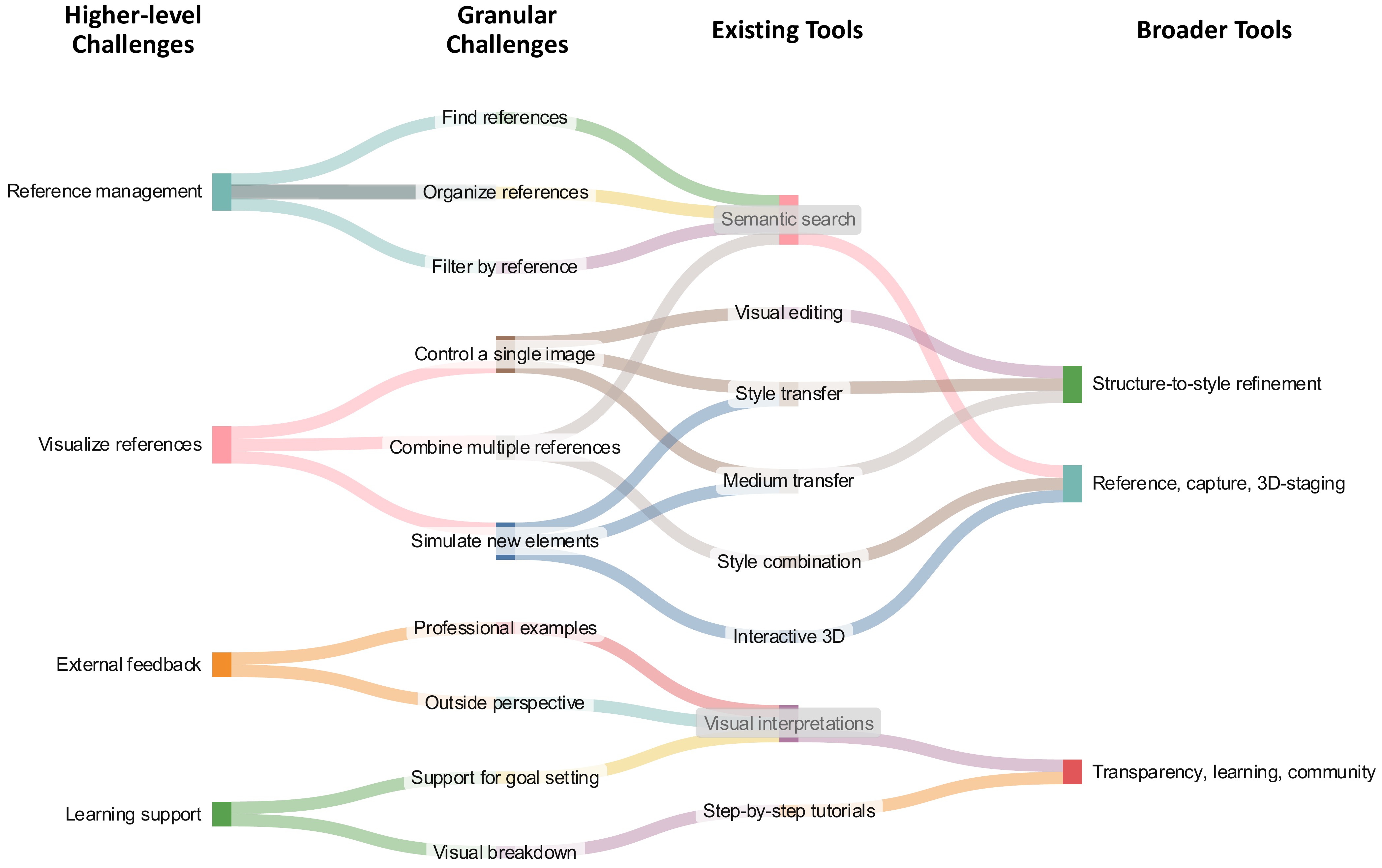}}
    \vspace{-0mm}
    \caption{\textbf{Mapping of higher-level and granular challenges, relevant existing tools, and broader generative tool categories} We find that users broadly face 4 higher-level challenges: (1) quicker and better access to references, (2) visualizing combinations of references, (3) perspective and feedback from another artist, and (4) support to learn art techniques and styles, which expand into ten more granular challenges. Based on these needs, we developed 6 prototypes demonstrating the potential for AI tools to augment creative workflows.}
    \label{fig:prototypes}
\end{figure}


Then, we invited participants to 30-min onboarding sessions to (1) share, discuss and reflect on the taxonomy of novice artist challenges and (2) introduce the six interfaces as prototype tools. 
The additional step of sharing Table \ref{tab:phase1-summary} intended to give participants opportunities to give feedback on and refine our taxonomy. 
The six interfaces were initially introduced at this stage so that participants can use prototypes on their own time, situated within their natural artistic workflows. 
To maximize comfort, we also warned participants that although our team would not explicitly collect data from their art pieces, the underlying models may not guarantee the same. 
We asked participants to reflect on how prototypes were helpful in the workflow, aspects that could be improved, and other concerns they had when interacting with each prototype. These reflections were written down prior to workshops. 
We then used these reflections to inform how we would facilitate discussions in the co-design workshops. Participants were compensated with \$20 USD for participation in the onboardings sessions as well as an additional \$20 for an estimated 30-45 minutes of interaction with our prototypes on their own time. 

\subsection{Phase 2 Co-design Methods}
\label{codesign} 


\subsubsection{Co-Design Workshops with Novice Artists.}
To further investigate potential tensions when implementing tools to address the four challenges identified, we conducted four workshops with 3–4 participants each --- recruited first from our interview pool, and later through social media postings and snowball sampling via referrals (see Table \ref{tab:workershop-demographics}). The workshops were designed to create space for artists to converse with one another, compare experiences, and collectively reflect on the promises and complications of emerging AI tools. This approach mirrors prior CHI work that uses workshop-based co-creation around early prototypes to envision and critique creative tools ~\cite{muller1993participatory}. Each session began with participants reflecting on six prototype concepts. We asked them to discuss challenges that might emerge when using such tools in their own practices, as well as opportunities for improvement. After this exercise, we introduced participants to examples of state-of-the-art tools identified through a survey of recent HCI and creativity support research.
These examples were not intended to be comprehensive, but rather to highlight representative directions in current research and to prompt further reflection on potential developments and frictions. Finally, throughout the discussions, we encouraged participants to consider the broader ethical and practical implications of generative AI systems in art-making --- asking them to reflect on risks that may arise for novices and communities, as well as to share any recommendations for how such systems could better support creative growth while mitigating unintended consequences. Participants were compensated with \$40 each for participation in co-design workshops.

\subsection{Positionality Statement}
\label{positionality}

This research was conducted by a multidisciplinary team of researchers based in the United States, with backgrounds in human-computer interaction (HCI), software engineering (SE) and machine learning (ML) across five institutions.
Our epistemologies --- as well as geographic and institutional positioning --- shaped the scope and framing of this work, including access to participants and assumptions around digital infrastructure, language, and AI tool availability. 
Several team members have prior experience in visual art and related creative fields, informing our interpretations of participants' accounts. 
However, we take a critical and reflexive approach throughout the process
to continuously sensitize to potential downstream impacts and ethical implications of this approach towards novice artists and the broader artistic community, especially regarding issues of authorship, labor, and creative autonomy. 
We acknowledge how our dual roles---as both system designers and analysts---position us in complex relations to the communities we study. Throughout this work, we aimed to center the lived experiences and values of novice artists, while reflecting critically on how our own positionalities shape the tools and narratives we construct and introduce, so as to avoid unintentionally substituting our voices over those of the artists, whom we aim to serve. Finally, we recognize that our identity as U.S.-based researchers limits our ability to fully capture the diverse and global artistic practices across experience levels and cultural contexts. 

\section{Phase 1 Findings} 

\subsection{Novice Visual Artists' Early-Stage Practices and Broader Contexts}
We report specific practices that novice visual artists currently engage in during early stages of their creative workflows.
\subsubsection{Motivations \& Communities for Art-Making.}
Compared to professional or outcome-driven goals, novice participants were more motivated by (1) \textbf{intrinsic enjoyment} and (2) \textbf{interactions with others} that provided support and sources of inspiration. For instance, P6 described how \textit{``a lot of the art that I create on a slightly more frequent basis \dots is for journaling purposes''}, while P4 perceived art-making as \textit{``a very solo experience \dots  [that] I do alone to connect with myself and make something that makes me feel better about the world''} (P4). Beyond reflections, P10 related the calming benefits of art-making: \textit{``I really enjoy drawing old toys. \dots It['s] just a relaxing thing for me''} (P10). For others, motivation was shaped by moments of external recognition or affirmation, with several novices crediting teachers, friends, or family members who encouraged them to create art. P10 recalled a class where the teacher
\textit{``ask[ed] me there}, `oh my gosh, are you an artist? This looks fantastic!' \dots \textit{So I started looking at Tiktok videos of it, and I was able to start painting with it''}. Several took up visual art after seeing other artist's work. Such experiences made novice artists like P12 \textit{``think of something that I've never thought of before. \dots and [then] I feel creative''} (P12). Together, these findings highlight how novice artists' motivation emerges from a blend of intrinsic enjoyment, relational encouragement, and a desire to respond creatively to the world around. 

Participants also gather input from people in their lives and sought broader ties with artists and communities. P12 asked friends \textit{``what's something [that] is not right about this image? \dots [Because if] the eye is in the wrong position, for example, I just can't see it''}. For others, being around other artists and their work was a beneficial source of inspiration (e.g., P4). In one interesting case, P11 enjoyed drawing with \textit{Art Fight}\footnote{https://artfight.net/} --- a platform bringing artists online together to create different versions of each other's work, which P11 described as \textit{``a drawing event that happens in July. Everybody joins a team, and you draw other people's characters, and you get points for it \dots it's the game''}. 
Other important sites of connection included art education (P5 and P9), online platforms (P2, P7), and informal peer networks (P4, P8). 

Crucially, while some participants described strong reliance on peer and community support, others reflected on the absence of such networks, leading to desires to partake in more robust or ongoing art communities.
P4 expressed how \textit{``I would love the opportunity to make art with other more serious artists, where there was some more trust built between us''}. 
This blend of community engagement and collaborative learning highlights nuanced ways novices sought connection and validation, underscoring complex needs for connection, including a desire for inspiration from other artists' works, engaging with them to build relationships and provide support. 


\subsubsection{Digital/AI Tool Usage \& Ethical Considerations.}
Uneven access to digital tools (and self-taught approaches to learning them) deeply influenced novices' creative practices -- leaving them intimidated or under-equipped. 
Toward digital and AI-supported tools, participants expressed a mix of (1) \textbf{intimidation} by unfamiliar tools and (2) \textbf{excitement} to embrace advanced digital and AI features but also (3) \textbf{nuanced views} on the ethics of generative AI tools.  
For some, even non-AI-based digital tools felt inaccessible and intimidating: \textit{``I'm not good with computers, so I don't really mess with [digital tools]''} (P9). Other participants expressed excitement about the practical upsides of AI, looking forward to uses such as AI reducing the burden of manual browsing to find inspiration and allowing faster and easier manipulation of references (P13).

However, many remained wary of the impact of using generative AI tools on creative vision, professional identity, and the integrity of the artistic community. 
P6 described an instance of a generated image reshaping their mental vision of a character in unintended ways, where the output \textit{``is completely inaccurate to how I first imagined her. She had a much rounder face. \dots I still have this mental image [of the character]\dots I'm sad that I didn't articulate that myself first before being biased by [the generated image].''} Others added that generated content often appeared uniform, overly polished, or illogical for learning or inspiration, with anatomical and compositional errors that could mislead practice (P13, P11). Several worried that AI encourages cultural sameness that devalues original work (P10). Concerns about professional identity centered on authorship and livelihoods. Participants cited the lack of ethical compensation models and the use of copyrighted material, and emphasized authenticity as a core value of the artist role. Some viewed AI generated art sold for profit as undercutting traditional income streams (P8, P11). 

Finally, participants raised issues of community integrity and infrastructure. Some wanted clear labeling to avoid inadvertently using AI generated references in their process (P13), others expressed discomfort with environmental costs and systems that operate without permission or compensation, making it hard to imagine adopting these tools: 
\begin{quote}
    \textit{``Every AI image that's generated is really environmentally destructive and that sucks. There's no way around that. There's no way to make this system better that prevents the environmental destruction of every single generated crop. And I struggle with that.''} (P4)
\end{quote}
Overall, participants weighed uneven access and limited familiarity against excitement about practical capabilities, while holding nuanced ethical positions on authorship, compensation, environmental impact, and community integrity.


\begin{table}[h]
\centering
\small
\setlength{\tabcolsep}{6pt} 
\begin{tabular}{p{6.0cm} p{4.0cm} p{3.5cm}}
\toprule
\textbf{Challenges} & \textbf{Suggestions} & \textbf{Relevant Existing Tools}  \\
\midrule

\textcolor{purple}{\textbf{\ref{refs}}} Quicker and Better Access to References &
\begin{itemize}[left=0pt,labelsep=0.5em,itemsep=0pt,topsep=0pt]
  \item[C1.1] Find references
  \item[C1.2] Organize references
  \item[C1.3] Filter by reference
\end{itemize} &
\begin{itemize}[left=0pt,labelsep=0.5em,itemsep=0pt,topsep=0pt]
  \item Semantic search
\end{itemize} \\

\midrule

\textcolor{purple}{\textbf{\ref{visualize}}} Visualizing Combinations of References &
\begin{itemize}[left=0pt,labelsep=0.5em,itemsep=0pt,topsep=0pt]
  \item[C2.1] Control a single image
  \item[C2.2] Combine multiple references
  \item[C2.3] Simulate new elements
\end{itemize} &
\begin{itemize}[left=0pt,labelsep=0.5em,itemsep=0pt,topsep=0pt]
  \item Visual editing
  \item Style transfer
  \item Medium transfer
  \item Style combination
  \item Interactive 3D
\end{itemize} \\

\midrule

\textcolor{purple}{\textbf{\ref{feedback}}} Feedback from Another Artist's Perspective &
\begin{itemize}[left=0pt,labelsep=0.5em,itemsep=0pt,topsep=0pt]
  \item[C3.1] Professional examples
  \item[C3.2] Outside perspective
\end{itemize} &
\begin{itemize}[left=0pt,labelsep=0.5em,itemsep=0pt,topsep=0pt]
  \item Visual interpretations
\end{itemize} \\

\midrule

\textcolor{purple}{\textbf{\ref{support}}} Support to Learn Art Techniques and Styles &
\begin{itemize}[left=0pt,labelsep=0.5em,itemsep=0pt,topsep=0pt]
  \item[C4.1] Support for goal setting
  \item[C4.2] Visual breakdown
\end{itemize} &
\begin{itemize}[left=0pt,labelsep=0.5em,itemsep=0pt,topsep=0pt]
  \item Visual interpretations
  \item Step-by-step tutorials
\end{itemize} \\

\bottomrule
\end{tabular}
\caption{Summary of Phase 1 findings. }
\label{tab:phase1-summary}
\end{table}

\textbf{\textit{Fig 1. Initial Taxonomy after Phase 1 Interviews}} 

\subsection{Areas for Support in the Early-Stages} 


Below, we present four key challenges: (1) 
\textbf{quicker and better access to references}, (2) \textbf{visualizing combinations of references}, (3) \textbf{perspective and feedback from another artist}, and (4) \textbf{support to learn art techniques and styles}. For each, we describe complex difficulties novice artists experienced and present examples of workarounds some used to adapt current tools or practices to fit their needs. 
Next, we follow-up with suggestions participants provided for ways to be supported in artmaking. 
We conclude with considerations for potential tool development, including envisioned challenges and suggested methods for mitigating and preventing potential harms or misuses. 

\subsubsection{Quicker and Better Access to References.} \label{refs}
Participants frequently struggled to find reference images that match their creative intent, reporting current tools as frustratingly limited in surfacing relevant or high-quality material.
Several participants regularly search references on Google Image Search (P1, 5, 12), Pinterest (P1, 2, 5, 9, 11, 13), and social media (P5, 7, 9, 11, 12, 13). P13 reflected how, as a novice artist \textit{``I have a hard time drawing from imagination, and so \dots I go on Pinterest and look up pictures''} (P13). 
For others, references were important technical models for poses of objects, or specific human bodies and expressions (P4, P5, P6). 

The difficulty of finding references that aligned with creative intent posed a persistent challenge for participants, who highlighted needs for more natural and intuitive ways to gather references. P8 envisioned
    \textit{``an Instagram where you can change your preferences \dots There's check boxes of what you like \dots [so you can] pick the images \dots that's better than having [to search with] words, because sometimes words mean different things for other people.''} (P8).
As P5 explained: \textit{``the ideal kind of [tool would be] if I could just imagine a picture and then like it came to life''} (P5). Others wanted \textbf{fine-grained search filters for specific elements} like human poses or expressions (P5, P8). 
Artists also cared about \textit{filtering out} noise in current tools (4.1.2). For novices like P7, this lack of support caused enough cognitive load to prevent them from creating art altogether:
\begin{quote}
    \textit{``It's hard to extract the most important parts from [a reference image]. That leads to two problems. [One] is that sometimes I just don't like [to] paint certain scenes because I think it's too busy. \dots It takes me a long time to choose the right thing to paint. ''} - P7
\end{quote}

To manage these gaps, participants developed various workarounds to obtain desired references. Strategies included iteratively editing search terms on Google Search or Pinterest to manually refine attributes --- e.g., specific mediums (P4), objects (P2), realistic human emotions (P8), particular camera angles (P5) certain poses (P4, P5). 
Many artists also manually changed reference images, using Photoshop to combine different objects in their planned art (P5, P13); P9 combined real-life references with certain artistic styles (P12). Others directly edited and interacted with a single reference  --- P7 developed an app to blur certain details in reference photos, but expressed a desire for more complex tooling: \textit{``it would be really cool if a system could take in a picture and just \textbf{get the main essence of the of the scene, and get rid of all the unnecessary details}''} (P7).
Lastly, participants described the manual experience of scrolling through their references (e.g., search engine results, own photos) -- while some enjoyed revisiting all their reference images (P9), others found manual browsing of their reference library to be time-consuming (P9).

\subsubsection{Visualizing Combinations of References.} \label{visualize}
Another challenge participants faced was combining multiple references into a coherent image, describing difficulties with translating photographs, styles, or poses into a unified artistic vision. For example, P5 explained the challenges of combining reference images with existing tools like Adobe Photoshop:
\begin{quote}
    \textit{``[I] was using different images and combining them all together. It takes time to Photoshop \dots I can just take those images separately and paint or draw them as draw them together, even though they're separate. [However] I think it would be easier to combine them but again, it's just too much work.''} - P5
\end{quote}

Some used interactive references in place of multiple reference perspectives. For example, P1 explained that if they needed a reference \textit{``for human anatomy, I would personally go to download an app, like a medical app \dots because it's a 3d map. You can just rotate around the body \dots you can see all the muscles and how they work''} (P1). These examples illustrate how existing tools for combining or interacting with reference images often only supported fragments of participants' needs, leading them to use workarounds for fitting pieces together. While such strategies helped approximate visions, participants often required significant effort to interpret or manually blend references.

While P1 was resourceful in finding such references, other artists expressed desires for a dedicated and accessible tool. Some envisioned interactive \textbf{systems for manipulating poses, perspectives, or stylistic features in real time}. When reflecting on an art piece they completed, P6 expressed that \textit{``it would have been really useful to have a pose of that thing falling in a realistic style''} that shows a \textit{``puppet reaching into the air with that particular pose from the angle that I want''} (P6). P8 shared a similar need to manipulate a human-like puppet and \textit{``move the face to make it mad''} (P8). 
These suggestions highlight artists' desire for tools that not only provide static references but also allow them to manipulate and combine references dynamically.

\subsubsection{Perspective and Feedback from Another Artist.}
\label{feedback}

Artists often struggled with identifying errors or areas for improvement on their own -- an artifact of lacking \textbf{timely, supportive feedback that matched their artistic intentions}. They highlighted unique challenges in this phase --- e.g., difficulty redoing parts of a piece without completely starting over or managing tools: \textit{``something that could have helped me catch that this arm had kind of [from a] wonky perspective before it got so far that I already colored and shaded -- it would have saved me a lot of time.''} (P3). This reflects a broader challenge of how artists access feedback and whose perspectives they can rely on.

Several participants expressed a desire for feedback that is \textbf{context-aware, flexible, and intentional}. For many, feedback is most valuable when it comes from someone with technical expertise or an “artist’s eye”, who offer objective critique on aspects like balance, composition, or technique. P8 often sought input from her boyfriend, an art major with formal training, while P4 appreciated informal encouragement and more serious critique in group art settings. These artists found immense value in outside perspectives, especially when they had become too focused and needed fresh eyes to spot issues or opportunities. The desire for feedback also extended to compliments, encouragement, and casual observations from peers, especially in group settings or informal art gatherings. P4 explained that sometimes even disagreements with others were helpful: \textit{``you might not initially be in alignment with somebody. They might tell you something [about your art]. You're like, `okay, I could see that.' [You] try that and you [realize] `oh, actually, I love it''} (P4). Together, these accounts illustrate the importance of feedback on technical accuracy, motivation and perspective-taking, highlighting the multifaceted role of outside input in sustaining creative practice.

Participants imagined tools that could replicate the sensitivity of an ‘\textit{artist’s eye}’, offering feedback that was \textbf{flexible, contextualized, and tailored to their intentions}. For example, P11 wanted a tool that could help encourage and motivate them: \textit{``[When] I want to give up, the tool is like, `no, you can do it''} feeling that such a tool would \textit{``be so fun. It'd be so cute, too. I feel like people would like that''} (P11). Overall, in imagining these tools, participants emphasized the need for both constructive and uplifting feedback, surfacing opportunities for tools that provide the sharpness of an artist’s eye or support with reassurances.

\subsubsection{Support to Learn Art Techniques and Styles.}
\label{support}

Learning new techniques and styles on their own was a common among novices, with some finding it difficult to bridge the gap between observing others’ work and applying those methods in practice. This challenge was especially acute when artists attempted to adapt across mediums or translate from one stylistic tradition to another. For instance, P2 noted that \textit{``I'm not an art professional, so sometimes I would still struggle with techniques of painting certain things ... originally, I was learning traditional Chinese watercolor, and that's very different from Western watercolor''} (P2). To manage these difficulties, artists relied on practices such as copying interpretations of specific subjects when first learning (P2), recreating works through ``artist studies,'' following tutorials, or adapting pieces into different styles. Watching time-lapse or process videos (P11, P12) also helped demystify the effort behind finished pieces, making the process feel more approachable. Yet, translating observation into implementation was not always straightforward -- some expressed frustration when tutorials or advice were either too advanced or not grounded in the practical realities of their tools and materials (P2, P7). 

As such, participants emphasized the importance of resources that tailor to their current goals and level of experience, whether through step-by-step guidance, adaptive feedback, or examples that reveal works-in-progress rather than only finished pieces.
P7 described an ideal tool they would use:
\begin{quote}
    \textit{``ideally you have this tool that's watching over you while you paint, or while you are making art. It learns to provide you feedback at the right moment so that you're not being handheld the whole way, but, at the same time, you're not like getting lost. At the same time it figures out what [you're] good at, what you still need to spend some timeimproving on.''} - P7
\end{quote}
Such opportunities highlight a desire for resources that are not only instructive but also responsive to an artist’s goals and intentions.

\subsection{Interactive Co-design Probes} \label{probes}
\begin{figure}[tbp]
    \centering
    \resizebox{1\textwidth}{!}
    {\includegraphics{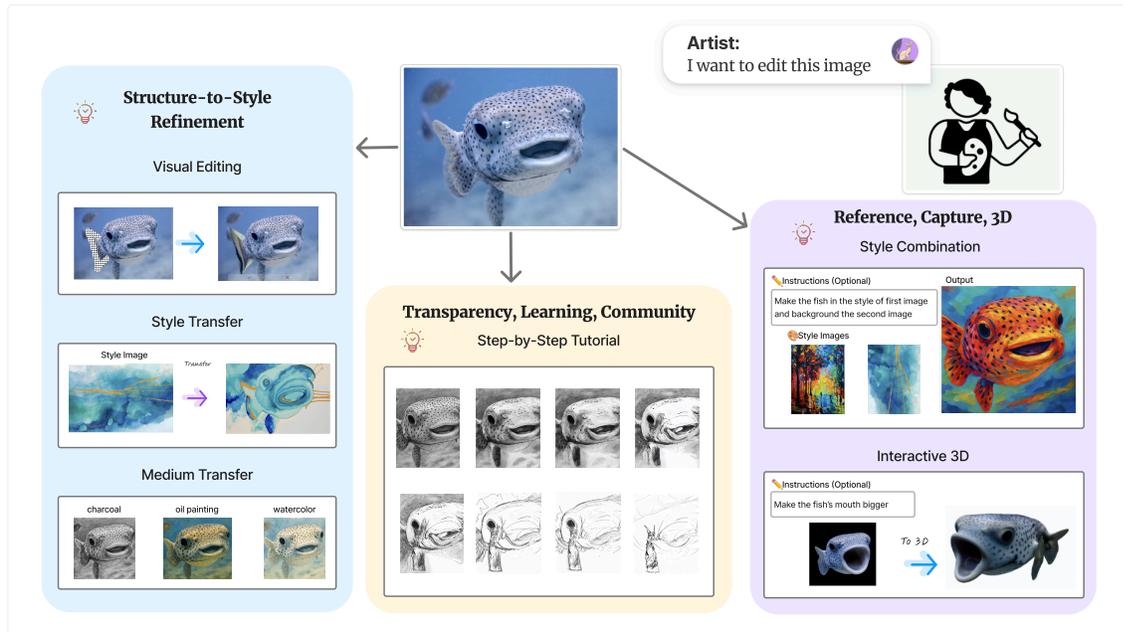}}
    \caption{\textbf{Development of Co-design Probes.} Six interactive prototypes were developed to address needs identified through user interviews and to explore AI-assisted design workflows, belonging to three broad categories. (1) Structure-to-style refinement tools help artists progressively convert initial ``structured'' forms (e.g., a basic shape, outline, style, or geometric layout) into more aesthetically complex or stylized versions, such as visual editing, style transfer, and medium transfer. (2) Reference, capture, and 3D-staging tools help search, store, and combine references to source material (e.g., style combination) in both 2D and multi-view 3D environments (e.g., interactive 3D creation from a single image). (3) Transparency, learning, and community-based tools help make the most salient parts of the input more transparent, such as step-by-step tutorials of visual content to help artists learn new techniques and styles.}
    \label{fig:prototypes}
\end{figure}

Based on novice artist challenges identified above, we developed interactive co-design probes that (1) responds to specific challenges while also (2) showcasing relevant state-of-the-art generative  AI capabilities. These tools fulfilled three broader categories: (1) reference, capture, 3D-staging, (2) structure-to-style refinement, and (3) transparency, learning and community. While we presented them in isolation, participants were free to use them in any combination with one another, potentially across multiple steps to achieve more desired effects.


\subsubsection{Structure-to-Style Refinement} tools progressively convert initial ``structured'' forms (e.g., basic shape, outline, style, or geometric layout) into more aesthetically complex or stylized versions. 

\paragraph{(i) \underline{Visual editing}. } Traditional image editing with generative models is largely text-driven, but artists often find text prompts insufficient for capturing precise edits. We leveraged MagicQuill~\citep{liu2025magicquill} to combines direct visual manipulation with generative editing, allowing users to sketch or mark directly on an existing image (need C2.1), reducing the burden of linguistic description while affording immediate, tangible control.

\paragraph{(ii) \underline{Style transfer}.}

To match artists' desires to observe other's work and translate others' stylistic tradition to their own (C2.2), 
we adapted FLUX Style Shaping\footnote{https://huggingface.co/spaces/multimodalart/flux-style-shaping}, a recent advance in generative editing based on ~\citep{flux} to enable flexible restyling while preserving image fidelity. 
Users provide (1) a \textit{structure image} defining the underlying composition, and a (2) \textit{reference image} to apply style onto. 
The system further contain controls to adjust balance between semantic preservation and stylistic strength.

\paragraph{(iii) \underline{Medium transfer}.}
Also based on C2.2, artists often wonder how a reference image might appear in a different medium --- e.g., \textit{a cat in oil paint re-rendered as a pencil sketch or watercolor}. 
We built an interface that used the GPT-image-1 editing model to transform user's image into their target medium.

\subsubsection{Reference, Capture, and 3D-staging.} 
These tools help search, store, and manage references to source material like images, videos, texts, or styles.
In addition to 2D source material and references, some tools in this category also extend to multi-view 3D environments.

\paragraph{(i) \underline{Style Combination}.}
Artists often draw on multiple references simultaneously, blending objects, motifs, or styles to spark unexpected ideas. A recurring challenge (C2.2) remains around merging elements into a single coherent composition — e.g., \textit{a realistic panda rendered in the aesthetic of a traditional Gyotaku fish print}.
Using GPT-Image-1~\cite{gpt4v}, we allowed artists to perform \textit{multi-reference combination} with a \textit{structure image} that defines the overall layout. 

\paragraph{(ii) \underline{Interactive 3D editing}.}
To combine 2D editing with 3D reconstruction, we again leveraged GPT-Image-1 to enables high-quality edits directly onto a 2D rendering. The modified image is then passed through Trellis~\citep{trellis}, to generate a corresponding 3D asset -- meeting C2.1.

\subsubsection{Transparency, Learning, and Community.} The final set help the creative and collaborative process among groups of human artists by making salient parts of inputs more transparent 
or summarizing the image using its key attributes. 

\paragraph{(i) \underline{Step-by-step tutorial}.} 
To support learning art technique and style from existing work
(C4.2), we built upon PaintsUndo\footnote{https://github.com/lllyasviel/Paints-UNDO} to prototype the generation of process-oriented tutorials videos from a static image to visualize intermediate steps of art making.


\section{Phase 2 (Workshop) Findings}

Below we report participants' reflections on interactions with our prototypes and existing capabilities of relevant tools. We then describe new capabilities that participants surfaced through the workshops and the ways these expand the design space. Finally, we synthesize cross-cutting design tensions that span the emerging taxonomy of capabilities, highlighting trade-offs that matter for future design and deployment.

\subsection{Reference, Capture, 3D-Staging} \label{reference}

\subsubsection{Semantic Search and Reference Organization.}
\label{semantic}

Participants sought tools for semantic search over large image and video sets, returning per frame matches and automatically organizing and deduplicating reference material~\cite{VSC, lin2014visual, OrganizingPhoto}.
Several participants described ideal use cases for tools that \textbf{search through videos}~\cite{GenQuery, VSC}, expressing desires to use them for the early stages of visual art-making. W10 explained potential benefits of performing an \textit{``\textbf{object search}''} on full videos like they have done for singular images. Additionally, W11 elaborated that a tool such as \textit{Textual Search Video Frame} ~\cite{lin2014visual} could help them do real-time multi-person tracking for interactive art projects that involve body movement. In short, participants described video-level search and frame-level retrieval as scaffolding for early ideation and blocking, which reduces the time spent hunting for references and enables faster iteration on pose and motion.

For a tool such as \textit{Organizing Image Collection}~\cite{OrganizingPhoto}, participants such as W13 expressed interest but noted that their personal photo libraries were not large enough to benefit: \textit{``if it were like a better Google of like finding references, that would be great. But if it were out of my own references, I just don't think I take that many photos''}. Instead, W13 explained that they wanted these functions for the reference images they save from Google Search or other sources online instead. 
Together, the findings suggest that semantic search and \textbf{automatic organization of reference libraries} would materially lower the cost of finding poses and motions during early stages for artists.

\subsubsection{Pose and Camera Control with On-demand Reference Capture.}
\label{pose}
Participants prioritized fine-grained pose and camera control, with on-demand reference capture and structure-preserving remix. They used Interactive 3D to explore complete angles and perspectives while comparing it to model-driven pose and view edits in prior tools~\cite{wu2025qwen}. Reflecting on their experiences with our \textit{Interactive3D} prototype, artists described 3D models as helpful for \textbf{viewing a subject from a specific perspective} (W11, 13). However, they also raised concerns about fidelity: \textit{``messed up anatomy, or [other] messed up things that are not true reality---I'm afraid of that lessening my knowledge of the real world''}. In short, participants valued \textbf{direct camera and pose control} in 3D as a dependable way to generate accurate references with trustworthy anatomical fidelity.

Beyond seeing its value for perspective referencing, participants also proposed concrete improvements to \textit{Interactive3D}. 
W1 wanted to change skin textures to make a model look \textit{``younger or older, or rougher or smoother''}, W2 desired fine-grained control of a model (e.g., manipulating precise finger and wrist poses) while W5 suggested integration with existing workflows (e.g., model download as an OBJ file). Taken together, these point to complementary workflows: (1) \textbf{study mode} with deep control and export for detailed reference building, and (2) \textbf{quick capture mode} that limits controls for speed and focus.

Other state-of-the-art tools (e.g., \textit{Qwen Image}) also offered useful features such as changing poses and perspectives~\cite{wu2025qwen}. For example, W9 valued support to change perspectives in 2D art because it shows new angles quickly. In general, participants prioritized \textbf{speed and convenience} for early ideation stages when they needed rapid alternative views without manual setup. In short, model-driven pose and view edits offered fast suggestions for exploration, while \textit{Interactive3D} provided grounded control for precise reference generation.

\subsubsection{3D-aware Editing of Complex Scene Changes.}
\label{aware}
Our analysis shows that artists sought tools capable of 3D-aware editing to change scenes in references. Participants reflected on use cases of our \textit{Interactive3D} prototype for \textbf{editing complex backgrounds} as well as state-of-the art tools such as \textit{VoxHammer}~\cite{li2025voxhammer}. For example, W10 explained that \textit{``if you're drawing a manga, and you need the city to be your background, then you can have a 3D-model and trace in the background''}. Moreover, W11 emphasized that that such 3D models for references are most useful when they \textit{``cannot picture the scene in [their] mind and [they cannot find the real scene in the world''}. These reflections position 3D-aware background editing as beneficial to scaffold early composition and scene continuity in cases where references may otherwise be less accessible to participants. 

Additionally, participants provided suggestions for improving 3D editing tools to fit their workflows. Some felt overwhelmed by the capabilities of 3D editing, mainly wanting the capability to make smaller changes. W8 explained that \textit{``I haven't used Blender. I haven't used ZBrush. So would find something like 3D editing of a model to be difficult if it's a whole scene''}. W12 expressed a similar desire to apply such tools for simplistic changes, such as in lighting a scene. In contrast, W11 wanted specific controls when editing such as the ability to edit a specific part of a scene while being able to lock the rest of the model. These concerns suggest benefits in gentle entry points and per object, region based controls that favor small, reliable edits like lighting or camera tweaks over full scene overhauls. 

\subsection{Structure-to-Style Refinement}
\label{s2s}

\subsubsection{Layered and Simplified Decomposition with Focus Control.}
\label{decompose}

Participants also expressed interest in tools for breaking references into meaningful layers and controlling what to focus on: layered, structure preserving simplification with simple toggles for focus~\cite{swiftsketch, neutralstroke, wu2025qwen}. 
For example, W9 highlighted a blockout view as especially useful for assisting them in reading landscape forms ~\cite{wu2025qwen}. Others noted the use of \textit{SwiftSketch} for reducing complex forms such as robots into simpler blocks (W4 and W10). Overall, these accounts point to imagined tool uses for layered decomposition, supporting exploration in early-stages.

We also surfaced pitfalls and changes needed for tools. Some disliked filter style simplifications that merely restyle images \textit{``when it asks it to be more simple it looks like a pencil sketch filter''} and qualitative prompts such as \textit{simple} that do not map to consistent levels of abstraction (W10). These participants asked for progressive breakdowns of references with features such as \textbf{structure-preserving edits, automatic pose abstractions}, and numeric controls such as \textbf{stroke count} (W10). Across layers, several participants also expressed desires to \textbf{mute or spotlight subjects}, and \textbf{plan major forms and composition} quickly (W8, W9, and W10)~\cite{neutralstroke}. Moreover, W8 wanted a tool that took a values-first view, \textit{``think[ing] in areas of dark and light \dots rather than objects,''} run alongside object-aware layers to avoid color filter distraction. These needs point to the potential of tools that can help participants break references into meaningful layers and control focus within references.

\subsubsection{Detailed Zoom, Pattern Extraction, and Transparent Adjustments.}
\label{zoom}
Participants wanted crisp zoom, resolution enhancement, and texture inspection that let them zoom into a style, isolate textures, patterns, and palettes, and make controlled variations with transparent local and global controls, including single-attribute adjustments. 

When testing our \textit{Visual Editing} tool, they praised the non-destructive history for safe exploration before touching the real piece, because it let them “trace your steps back or forward” (W3). They were frustrated when style controls felt vague or cosmetic. The ``pencil sketch'' option did not behave like real pencil and lines felt generic, and the tool handled photo fixes or collage effects better than preserving a specific drawing style (W8).
Edits sometimes ignored source style or color requests, for example asking for ``pink pastel'' produced neon or dark tones, and overall the tool often drifted into a different style than the original (W4). Participants also asked for fine control over faces and materials, indicating interests in structure-preserving when applying local changes to an image. They also noted upon zooming-in and extracting details for changes, tools should be intent-aware and edge-aware that goes beyond rigid selections and proactively clarifies ambiguous requests. For example, when a user says ``change shirt to hoodie,'' the tool should ask whether to replace the garment, add a graphic, recolor it, or adjust the fit (W10). 

Other tools helped participants round out this capability. W4 preferred outfit-focused detail zoom from \textit{Qwen-Image} for recreating historical clothing, asking for true pattern extraction so users can capture motifs for cosplay or pattern making. W8 praised \textit{GenQuery}-style controls that vary a single element inside a scene -- they also asked for a simple adjustment slider to study and change styles with clear controlled variation. W12 valued resolution upscaling to turn low-resolution assets into high-resolution references, as long as structure stays intact and details do not get made up. 

\subsubsection{Controllable Style Analysis, Transfer, and Combination.}
\label{style}
When learning a new style, participants asked for tools that decompose style parts such as technique, color, form, and stroke, then let them mix or remap those parts with live previews. They also desired to preserve structure when applying style transformations so edges and keypoints stay fixed, as well as the ability to learn strokes from simple outlines and rule sets that state what may change or stay put. 

With \textit{Medium Transfer} and \textit{Style Transfer}, participants reported how the tool helped them rethink linework and color choices and refine details. W1 used multiple examples to make the system learn ``key aspects'' of their style and subsequently embed them in a new image -- this process also surfaced insights about their own style, although current outputs still felt too generic. Others observed that the model often remapped colors without altering structure in useful ways but wanted stronger structure control. W12 asked for a \textbf{structure-preserving mode} and \textbf{style isolation} to exclude frames or backgrounds during style analysis, plus fine controls for colorfulness, brightness, and local composition so individual style components can be adjusted without affecting the whole image. 

For \textit{Style Combination}, participants considered it a way to learn new styles through controlled change and asked for two exploration paths. One is a \textbf{spread of quick variations for side-by-side comparison} when studying a style. The other is a \textbf{single precise output} when the target look is already clear. Participants additionally suggested readable diagnostics that turn results into study guides, such as palette and brush breakdowns tied to each output.

When manipulating styles, participants emphasized region-aware conrol to assign one style to the background and another to the subject, and guardrails for content exclusion and frame-to-frame consistency when fusing styles. For example, W4 mapped a pastel style onto a 3D character and got clean line art and a coherent palette. Image-to-image conversion kept pose for anatomy study (W10). A line-art-to-fantasy conversion captured the target aesthetic and would benefit from annotated palettes and brushes for analysis (W9). Participants used combinations to generate multiple exampels to reverse-engineer what works across outputs (W1), and noted that fixed structure with varying style is especially inspiring for study and practice (W11, W12). 


\subsection{Transparency, Learning, Community}

\subsubsection{Explainable and Transparent Learning Workflows.}
\label{transparent}

Participants expressed interests in tools that pair \textbf{in tool guidance} with \textbf{shareable provenance} so they can learn how to pick strong references, understand why results look the way they, and quickly remix or refine with better prompts or sketches. They want clear instructions, guided prompts, and step by step tutorials tied to exact steps, models, and settings that can be viewed, studied, and rerun.

Experiences with our \textit{Step-by-Step Tutorial} surfaced what to keep and what to change. W4 valued a staged breakdown that moves from flats to simple shading to color and texture, since it reveals how effects like gloss are built rather than guessed from a finished image. W1 asked for tutorials that specify the exact actions needed to reproduce another artist's coloring and noted that stimulating human mistakes is distracting in a tutorial. Others like W7 and W8 found that the current tool behavior too much like tracing or reveal by parts rather than "a natural process that humans will draw" that starts with gesture, construction, form and shading, then detailing (W12). They asked for a true \textbf{draft to final pipeline} that shows how features and color layers emerge in a natural order.

Participants asked for an explainable and tranpsarent learning flow that shows real making steps to understand cause and effect relationships. W8 said current tutorials like \textit{PaintsAlter} and \textit{Step-by-Step Tutorial} often jump from canvas to finished scene with limited intermediate steps; instead they want to see outlines appear, colors laid down in order, and details added gradually so they can answer ``what order did they lay the colors down.'' They also wanted a single place to pin references and move step by step, describing a workflow of ``I want to draw this, show me how.'' W4 emphasized tools should make intent legible and repeatable. Having a \textbf{scaffolded input box} improves results by letting participants specify exactly what they want, and in 3D they can apply the same logic to learn from pose and structure. Taken together, participants' requests point to guided prompts and staged tutorials that mirror human practice so artists can learn and stay in control.

\subsubsection{Community Spaces and Feedback Loops.}
\label{community}
Participants asked for \textbf{lightweight community spaces} that support sharing references, running remix challenges, and reusing each other’s \textbf{plug-and-play pipelines}. They want \textbf{fast goal-aware feedback} from AI, optional human critique, and clear privacy and consent controls upon joining a community. Without a prototype, they reacted to the concept and pointed to gaps in current tools. W8 described a simple social layer where artists post references, view many remixes of the same prompt or image, and browser the exact steps behind a piece. They envisioned a \textbf{Hugging Face for artists} where creators publish pipelines with models, steps, settings, and notebooks that others can reuse or adapt. Capturing how pipelines are used would help artists and surface lesser know models that are hard to discover.

Participants highlighted adjacent needs that current platforms do not meet. W4 asked for an \textbf{up-to-date meme library} with searchable templates, recency signals, and links to original post since general search often returns stale content. They wanted the same for reference search so current and culturally relevant formats are easy to find and import straight into a working file or pipeline. W8 also pointed to \textbf{asset-first sharing} where people publish building blocks and explain how they were used to strengthen learning and reproducibility.

For critique, participants want a hybrid loop that respects intent. When shown the idea of \textit{CognArt} with analysis trends and feedback based on breakdown elements, W10 still favored real human critique: “If I have access to a real human, I would prefer a real human.” W11 proposed \textbf{role-played AI critics}, such as a gallery curator, animator, or hobbyist, to broaden perspectives when those voices are not available to artists on demand. W12 recommended combining both: use AI for fast, professional suggestions when improving or monetizing work, seek human attention when audience building matters, and skip feedback entirely when creating just for fun. They noted that high quality human critique is hard to get, which is where \textbf{structured objective AI guidance} can help, with the option to schedule human reviews when needed.


\subsection{Design Tensions}










In this section, we highlight tensions that emerged through participants' reflections on capabilities in co-design workshops. 

\subsubsection{Assistance vs Independence.} 
A major tension emerged around how much assistance models should provide while still preserving novice artists agency and workflow. Participants welcomed tools that could teach and scaffold within the early-stages of their workflow, but insisted that such guidance should not steer their work. W8 wanted live \textit{``what this looks like''} matches based on art history and pop-culture to build taste and test how others might read their work, but warned that real-time labeling risks boxing them in byimporting biased and clichéd references, discouraging originality, and breaking creative flow (W8). W8 suggested offering multiple procedural paths with a simple timeline to switch among them, prompts such as \textit{``what else could go here?''}) to invite abstraction, and light intent checking (\textit{``here’s how I read your prompt, here is how my model works''}) so agency remains with the artist(W8).

Participants also clarified where feedback belongs and how to deliver it without taking over. For example, W13 advised that tools should not \textit{``take the fun out of art''}. W10 proposed a division of labor for feedback:
\begin{quote}
    \textit{``since artists want feedback yet fear AI feedback being generic and nonconstructive, perhaps the ideal case would be to have AI be good at giving feedback for \textbf{ground-truth domains} with expert rubrics (e.g., anatomy) but avoid \textbf{prescriptive judgments} for subjective art and self expression (e.g., when helping with style or color palettes without standard definitions)''}
\end{quote}
Non destructive options such as temporary overlays or side-by-side-comparisons can keep artists original work intact while building on suggestions. Overall participants advocated for tooling that assisted them rather than make decisions for them, providing adjustable guidance under their control that improves craft while preserving flow. 

\subsubsection{Unexpected Surprise versus Accuracy.} 
Another concern was with how much to tools should welcome surprise versus enforce accuracy. For many participants, serendipity was valuable for inspiration. W1 accidentally swapped the order of the style combination input, but was surprised to be happpy with the output, noting they liked the \textit{``hard shadows''} and \textit{``fuzzy neon highlights''}. Further, W2 noted that failed outputs may be \textit{``more helpful than the successful ones''} since it could be weirder or more unexpected in a way that provides them with inspiration. In these cases, we found that random and vague outputs were useful for open-ended brainstorming especially when borrowing elements rather than replicating them. In contrast, for committed work, participants emphasized the importance of fidelity and control. In this way, W2 cautioned that current systems are better for ideation than for early sketches. W1 and W3 expressed a more optimistic option, citing that tools such as our \textit{Interactive3D}prototype could be helpful for pose changes of references if only the lighting, shadows, and model proportions were close to real life. 

Moreover, participants also weighted realism against the importance of pieces having a human touch to them. W7 explained how the generated image in style transfer \textit{``looks so good''} but also felt like \textit{``AI art''}. W2 wanted integrated AI tools (in software like Procreate/Photoshop) that improved their workflow by fixing proportions and lighting, or refining their style, but without generic outputs that look \textit{``too AI''}. Overall, participants wanted controlled instances of surprise, in which tools could invite unexpected variation during exploration while preserving precise, non-generic and human-guided results as their work moved toward completion. 



\subsubsection{Values versus Concerns of AI.} 
There was also tension between artists, who despite seeing the value of AI as tools, are also sensitive to their potential in exploiting artists like themselves. Participants saw value in tools that took advantage of AI capabilities, but hestitated to trust them. W3 found it difficult to use AI tools to help improve their art because the even the act of uploading their art means that AI could plagiarize it. Moreover, W1 said that they would not unless contributing artists had granted permission for the use of their work. W5 further noted that they would feel comfortable only once artists were fairly compensated as well. As such, we found that participants tied trust to practices of consent and transparency, advocating for AI that was trained on public domain or opt-in licensed data and payment to artists. 

Furthermore, participants wanted to be able to teach AI systems without copying the output. The reasoning behind this stemmed from a view of style as personal characteristic that some participants compared to handwriting (e.g., W10). W1 argued that \textit{``drawing a reference generated exactly from AI is the same as copying someone else's art and that you shouldn't be able to claim that as your own work, but you can copy as more of a learning process''} (W1). W8 advocated for turning tools into learning aids to help artists draw their own icons, for example, rather than copying the words of others (W8). Overall, participants strongly advocated for consent-based, transparent tools that could help novice artists such as themselves build skills and provide guidance while avoiding appropriation. 



\section{Discussion}

\subsection{Supporting Artistic Skill-building via Enhanced Agency \& Control}


\subsubsection{Individual-Level Support}
Our analysis points to concrete opportunities for individual-level support of novice artists in the early stage of art making. Novice artists want support to create in the early stages with confidence while preserving agency and authorship. Prior work on agency, disclosure, and learning offers useful principles for early-stage assistance~\cite{shneiderman2022human, amershi2019guidelines}. There are clear opportunities to support art skill-building with capabilities that improve interaction and access to references. Useful directions include guided reference curation, similarity and diversity controls for exploring sets of references, sketch to reference matching, and lightweight prompts that teach search strategies as people work ~\cite{vygotsky1978mind, wood1976role, puntambekar2005tools}. Interfaces should foreground provenance and allow people to filter by source type, license, and stylistic attributes so that reference use is intentional and creditable ~\cite{mitchell2019model, gebru2021datasheets, bender2018data}. In this frame, the system may serve as a studio partner that accelerates exploration without overriding intent. Control is a central requirement for individual artists who interact with these tools. Participants wanted adjustable model influence, visibility into sources and transformations, and simple ways to accept, revise, or discard suggestions ~\cite{amershi2019guidelines, lubart2005can}. Designs should default to reversible edits, versioning, and private workspaces, with explicit choices to share or credit others. Such controls help novices learn with assistance while keeping authorship clear and decisions recoverable.

\subsubsection{Interaction-Level Support}
Our findings indicate that early-stage creativity is not only individual but also deeply social. Participants asked for spaces that make it easy to learn with and from others, frequently imagining a Hugging Face for artists that hosts references, process notes, prompts, and works in progress with clear credit ~\cite{ko2023large, wang2025aideation, everybody_sketch, PortraitSketch}. The design problem at this level is less about accelerating a single user’s search and more about building the social and technical infrastructure that turns many small contributions into shared knowledge ~\cite{magiccolor, yan2025imagereferencedsketchcolorization, trellis, wang2021screen2words, li2025voxhammer}. Distinct capabilities emerge in shared spaces. Discovery and matchmaking can connect novices to peers, mentors, and example workflows that fit their goals and skill levels ~\cite{chi2020automatic, chi2022synthesis}. Second, lineage-aware remixing can record how assets, prompts, and sketches relate, so that reuse is traceable and creditable across iterations ~\cite{lima, kawakami2024impact, foreground, giacomin2023intersection}. Community challenges and curated playlists of processes can function as shared curricula that scaffold progress from simple studies to more independent projects ~\cite{puntambekar2005tools}. In sum, interaction level support treats the system as a studio commons rather than only a generator of outputs. By investing in discovery, lineage, feedback, and governance, these spaces can turn individual experiments into collective learning while keeping consent, credit, and accountability legible.

\subsection{Policy Recommendations for ethical development/training AI/compensation}
Our analysis points to the importance of policy that guides the positive use of these tools so novice artists can build skills while avoiding exploitation. Prior work documents a history of harm to artists in AI development, including unconsented training~\cite{shi2023understanding, decolonial, giacomin2023intersection}, weakened attribution and compensation~\cite{kawakami2024impact, foreground}, and the straightening of expression through biased and homogenized outputs \cite{unstraighten}. We read our findings as prompts for community dialogue and governance rather than fixed prescriptions. Trust rests first on consent, compensation, and provenance. Participants linked willingness to use integrated features to permission and pay, and asked for visible signals about where data came from. As such, it may be beneficial to favor public domain or opt-in licensed data, compensate contributors, and make provenance and labeling visible across search, editing, and sharing so creators know what they are using. Clear source tracing and labeling also reduce the risk that novices inadvertently treat synthetic images as authoritative references.

Policies should also preserve agency while still offering help. Participants welcomed guidance that teaches and scaffolds, provided it does not steer the work or collapse stylistic diversity. Sensible defaults include opt-in guidance with visible intent checks, multiple procedural paths rather than a single route, and feedback that is confident where ground truth exists, such as anatomy, perspective, and lighting, while remaining non-prescriptive for subjective choices like style or palette. When tools produce references for pose, perspective, or lighting, accuracy safeguards and quality warnings are prudent so learners are not taught the wrong thing, especially in 3D reference workflows. Finally, participants imagined community infrastructure that supports learning without eroding integrity. Provenance-rich spaces that share processes and remixes, coupled with privacy and consent controls, are more likely to sustain participation and skill building~\cite{mitchell2019model, gebru2021datasheets}.  In sum, our findings motivate higher-level guidelines and accountability across development, use, and audit.

\subsection{Limitations \& Future Work}
Our qualitative approach enabled rich and nuanced accounts of early-stage art making, yet it does not represent the full diversity of novice artists. Future work should broaden coverage with larger surveys across regions and training backgrounds, and with comparative studies that include non-student and community artists. Additionally, self-report and reflection may not fully align with the actual practice of our participants, as is the limitation of the nature of our methodology that has been previously acknowledged~\cite{jansen2021exploring}. Participants reflected on ethical and policy questions specific to early-stage tools, yet it was not within the scope of our study to evaluate legal frameworks or institutional processes in depth. Future work should connect design proposals to analyses of consent, compensation, provenance requirements, and audit mechanisms in real organizational settings. Finally, while we utilized the interdisciplinary expertise of our research team to inform the development of our prototypes, it would be beneficial for future work to further examine the perspectives of commercial generative AI model practitioners to better understand the challenges in developing techniques such as semantic search. Accurate semantic search across multiple visual mediums remains a challenge in generative AI, although recent efforts in multimodal retrieval-augmented-generation~\citep{mei2025survey} are closing this gap. Other capabilities related to providing visual feedback and supportive learning communities to novice artists are also difficult, since the majority of generative models are critically lacking in social and embodied intelligence~\citep{mathur2024advancing}. Despite these challenges, we believe that our contributions can inspire the design of next-generation generative AI systems that support the creativity, learning, and agency of novice artists.
\section{Conclusion}
This project set out to understand how novice visual artists begin their creative workflows, what challenges they face in the early stage, and where emerging generative AI could offer support. Through formative interviews and co-design workshops, we observed that novices already rely on an ecosystem of tools for gathering references, sketching first ideas, and testing directions. Building on these practices, we distilled a set of empirically grounded capabilities that artists want from generative AI to scaffold exploration, spark variety, and help them learn without taking away creative control. At the same time, the study surfaced tensions that shape whether and how generative AI belongs in the early stage. Participants voiced concerns about authorship, credibility, and the risk that AI may limit personal style. They also worried about the opacity of training data and the absence of reliable credit and compensation for the creators whose work may inform model outputs. These findings suggest that design and policy must advance together. Systems should make provenance and attribution legible, allow artists to set boundaries on source use, and create space for human-guided iteration. Policy should address data governance and artist compensation mechanisms so that early-stage assistance does not come at the expense of artistic labor and trust. Our results point to promising opportunities for tools that support brainstorming, reference curation, and skill-building while preserving agency and voice. Future work may prototype and evaluate tools with capabilities for reference gathering, guided variation, and explanation of process. Continued research may also study their effects on learning and confidence over time. In sum, we find opportunities for generative AI to play a constructive role in early-stage creative workflows if it is designed for partnership rather than replacement and grounded in policies that respect artists and their contributions.

\bibliographystyle{ACM-Reference-Format}
\bibliography{_references}
\end{document}